\documentclass{article}
\def\Ref#1{(\ref{#1})}
\usepackage{amsmath}
\usepackage{amssymb}
\usepackage{cite}

\def\d{{\rm d}}
\newcommand{\A}{\operatorname{\mathcal{A}}}
\newcommand{\B}{\operatorname{\mathcal{B}}}
\newcommand{\BB}{\operatorname{\mathnormal{B}}}
\begin{document}
\begin{titlepage}
\noindent{\large\textbf{Perturbative calculation of one-point
functions of one-dimensional single-species reaction-diffusion
systems}}

\vskip 2 cm

\begin{center}{Mohammad~Khorrami{\footnote
{mamwad@mailaps.org}} \& Amir~Aghamohammadi{\footnote
{mohamadi@azzahra.ac.ir}}} \vskip 5 mm

\textit{  Department of Physics, Alzahra University,
             Tehran 19938-91167, Iran. }
\end{center}

\begin{abstract}
\noindent Perturbations around autonomous one-dimensional
single-species reaction-diffusion systems are investigated. It is
shown that the parameter space corresponding to the autonomous
systems is divided into two parts: In one part, the system is
stable against the perturbations, in the sense that largest
relaxation time of the one-point functions changes continuously
with perturbations. In the other part, however, the system is
unstable against perturbations, so that any small perturbation
drastically modifies the large-time behavior of the one-point
functions.
\end{abstract}
\begin{center} {\textbf{PACS numbers:}} 05.40.-a, 02.50.Ga

\end{center}
\end{titlepage}
\section{Introduction}
Reaction-diffusion systems, is a well-studied area. People have
studied reaction-diffusion systems, using analytical techniques,
approximation methods, and simulation. The approximation methods
may be different in different dimensions, as for example the mean
field techniques, good for high dimensions, generally do not give
correct results for low-dimensional systems. A large fraction of
analytical studies, belong to low-dimensional (specially
one-dimensional) systems, as solving low-dimensional systems
should in principle be easier.
\cite{ScR,ADHR,KPWH,HS1,PCG,HOS1,HOS2,AL,AKK,RK,RK2,AKK2,AM1}.

One of the reasons people want to find and solve exactly-solvable
systems, is that one can use perturbative methods to investigate
other systems, which are not exactly solvable but \textit{near}
some exactly-solvable systems. The term exactly-solvable have been
used with different meanings. For example, in \cite{AA},
\cite{RK3}, and \cite{RK4}, solvability (or integrability) means
that the $N$-particle conditional probabilities' S-matrix is
factorized into a product of 2-particle S-matrices; while in
\cite{BDb,BDb1,BDb2,BDb3,Mb,HH,AKA,KAA,MB,AAK}, solvability means
closedness of the evolution equation of the empty intervals (or
their generalization).

In \cite{GS}, a ten-parameter family of reaction-diffusion
processes was introduced for the systems among which, the
evolution equation of $n$-point functions contains only $n$- or
less- point functions. We call such systems autonomous. There, for
these models the average particle-number in each site was obtained
exactly. In \cite{AAMS,SAK}, this has been generalized to
multi-species systems and more-than-two-site interactions.

Among the important aspects of reaction-diffusion systems, is the
phase structure of the system. The static phase structure concerns
with the time-independent profiles of the system, while the
dynamical phase structure concerns with the evolution of the
system, specially its relaxation behavior. In
\cite{MA1,AM2,MAM,MA2}, the phase structures of some classes of
single- or multiple-species autonomous reaction-diffusion systems
have been investigated. These investigations were based on the
one-point functions of the systems.

In \cite{2point}, the two-point functions of autonomous
single-species translationally-invariant one-dimensional
reaction-diffusion systems were studied. The two-point function
for such systems was obtained, and it was shown that it exhibits a
non-trivial dynamical phase structure. The dynamical phase
structure of the system  was also  investigated.

In this article, we want to use perturbation to study systems
which are not exactly autonomous, but \textit{nearly} autonomous.
By this, it is meant that the rates of these systems are equal to
those of an autonomous system, plus a small perturbation. The
scheme of the paper is the following. In section 2, autonomous
systems are briefly introduced. In section 3, non-autonomous
perturbations around autonomous systems are considered, and their
effect on the evolution of one-point functions is investigated.
From this investigation, it turns out that some of the autonomous
systems are unstable with respect to perturbations, in the sense
that any small perturbation drastically modifies the large-time
relaxation of the one-point functions. Section 4 is devoted to a
concrete example.

\section{Autonomous systems and the evolution equations}
To fix notation, let's briefly introduce the autonomous systems.
Consider a one-dimensional periodic lattice, every point of which
either is empty or contains one particle. Let the lattice have
$L+1$ sites. The observables of such a system are the operators
$N_i^\alpha$, where $i$ with $1\leq i\leq L+1$ denotes the site
number, and $\alpha=0,1$ denotes the hole or the particle: $N_i^0$
is the hole (vacancy) number operator at site $i$, and $N_i^1$ is
the particle number operator at site $i$. One has obviously the
constraint
\begin{equation}\label{1}
s_\alpha N^\alpha_i=1,
\end{equation}
where ${\mathbf s}$ is a covector the components of which
($s_\alpha$'s) are all equal to one. The constraint \Ref{1},
simply says that every site is either occupied by one particle or
empty. A representation for these observables is
\begin{equation}\label{2}
N_i^\alpha:=\underbrace{1\otimes\cdots\otimes 1}_{i-1}\otimes
N^\alpha\otimes\underbrace{1\otimes\cdots\otimes 1}_{L+1-i},
\end{equation}
where $N^\alpha$ is a diagonal $2\times 2$ matrix the only nonzero
element of which is the $\alpha$'th diagonal element, and the
operators 1 in the above expression are also $2\times 2$ matrices.
It is seen that the constraint \Ref{1} can be written as
\begin{equation}\label{3}
{\mathbf s}\cdot{\mathbf N}=1,
\end{equation}
where ${\mathbf N}$ is a vector the components of which are
$N^\alpha$'s. The state of the system is characterized by a vector
\begin{equation}\label{4}
{\mathbf P}\in\underbrace{{\mathbb V}\otimes\cdots\otimes{\mathbb
V}}_{L+1},
\end{equation}
where ${\mathbb V}$ is a $2$-dimensional vector space. All the
elements of the vector ${\mathbf P}$ are nonnegative, and
\begin{equation}\label{5}
{\mathbf S}\cdot{\mathbf P}=1.
\end{equation}
Here ${\mathbf S}$ is the tensor-product of $L+1$ covectors
${\mathbf s}$.

As the eigenvalues of the number operators $N^\alpha_i$ are zero
or one (and hence these operators are idempotent), the most
general observable of such a system is the product of some of
these number operators, or a sum of such terms. Also, the
constraint \Ref{1} shows that the two components of ${\mathbf
N}_i$ are not independent. so, one can express any function of
${\mathbf N}_i$ in terms of
\begin{equation}\label{6}
n_i:={\mathbf a}\cdot{\mathbf N}_i,
\end{equation}
where ${\mathbf a}$ is an arbitrary covector not parallel to
${\mathbf s}$. Our aim is to study the evolution of the one-point
functions ($\langle n_i\rangle$'s).

The evolution of the state of the system is given by
\begin{equation}\label{7}
\dot{\mathbf P}={\mathcal H}\;{\mathbf P},
\end{equation}
where the Hamiltonian ${\mathcal H}$ is stochastic, by which it is
meant that its nondiagonal elements are nonnegative and
\begin{equation}\label{8}
{\mathbf S}\; {\mathcal H}=0.
\end{equation}
The interaction is nearest-neighbor, iff the Hamiltonian is of the
form
\begin{equation}\label{9}
{\mathcal H}=\sum_{i=1}^{L+1}H_{i,i+1},
\end{equation}
where
\begin{equation}\label{10}
H_{i,i+1}:=\underbrace{1\otimes\cdots\otimes 1}_{i-1}\otimes H
\otimes\underbrace{1\otimes\cdots\otimes 1}_{L-i}.
\end{equation}
(It has been assumed that the sites of the system are identical,
that is, the system is translation-invariant. Otherwise $H$ in the
right-hand side of \Ref{10} would depend on $i$.) The two-site
Hamiltonian $H$ is stochastic, that is, its non-diagonal elements
are nonnegative, and the sum of the elements of each of its
columns vanishes:
\begin{equation}\label{11}
({\mathbf s}\otimes{\mathbf s})H=0.
\end{equation}

Using
\begin{equation}\label{12}
{\mathbf s}\otimes{\mathbf s}({\mathbf a}\cdot{\mathbf
N})\otimes({\mathbf b}\cdot{\mathbf N})H=a_\alpha\, b_\beta\,
H^{\alpha\beta}{}_{\gamma\delta}{\mathbf s}\otimes{\mathbf
s}N^\gamma\otimes N^\delta,
\end{equation}
where $\mathbf a$ and $\mathbf b$ are arbitrary covectors, one can
write down the evolution equations of the one-, two-, or more-
point functions of $n_i$'s. The evolution equation for the
one-point function is
\begin{equation}\label{13}
\frac{\d}{\d t}\langle n_i\rangle=a_\alpha\,s_\beta\,
H^{\alpha\beta}{}_{\gamma\delta}\,\langle
N_i^\gamma\,N_{i+1}^\delta\rangle+s_\alpha\,a_\beta\,
H^{\alpha\beta}{}_{\gamma\delta}\,\langle
N_{i-1}^\gamma\,N_i^\delta\rangle.
\end{equation}
It is seen that the right-hand side of the above equation,
contains two-point functions. In fact, in the evolution equation
of $n$-point functions, there are generally up to $(n+1)$-point
functions. However, there are systems for them in the evolution
equation of $n$-point functions, only up to $n$-point functions
arise. These are the autonomous systems. For a system with the
Hamiltonian $H^0$ to be autonomous, following constraints hold
among the reaction rates nd their reaction
rates~\cite{GS,AAMS,SAK}.
\begin{equation}\label{14}
\sideset{^e}{^{0\alpha}{}_{\gamma\delta}}\A
=\sideset{^e_1}{^{0\alpha}{}_\gamma}\A\,s_\delta+
\sideset{^e_2}{^{0\alpha}{}_\delta}\A\,s_\gamma,
\end{equation}
where
\begin{align}\label{15}
\sideset{^1}{^{0\alpha}{}_{\gamma\delta}}{\A}:=&s_\beta\,
H^{0\alpha\beta}{}_{\gamma\delta}\nonumber\\
\sideset{^2}{^{0\alpha}{}_{\gamma\delta}}{\A}:=&s_\beta\,
H^{0\beta\alpha}{}_{\gamma\delta}.
\end{align}
It is not difficult to see that the constraints \Ref{14} are
equivalent to
\begin{equation}\label{16}
H^0\,{\mathbf u}\otimes{\mathbf u}=\lambda\,{\mathbf
u}\otimes{\mathbf u},
\end{equation}
where
\begin{equation}\label{17}
{\mathbf u}:=
  \begin{pmatrix}
    1\\
    -1
  \end{pmatrix},
\end{equation}
and it is obvious that
\begin{equation}\label{18}
{\mathbf s}\cdot{\mathbf u}=0.
\end{equation}

Now, consider an autonomous  system satisfying the constraints
\Ref{14} (or equivalently \Ref{16}), and take the vector ${\mathbf
v}$ satisfying
\begin{align}\label{19}
\left(\sum_{d,e=1}^2\sideset{^{d}_e}{^0}\A\right){\mathbf
v}=&0,\nonumber\\
{\mathbf s}\cdot{\mathbf v}&=1,
\end{align}
and the covector ${\mathbf a}$ such that
\begin{equation}\label{20}
{\mathbf a}\cdot{\mathbf u}=1,\qquad {\mathbf a}\cdot{\mathbf
v}=0,
\end{equation}
that is, the basis $\{{\mathbf a},{\mathbf s}\}$ is dual to
$\{{\mathbf u},{\mathbf v}\}$. In \cite{AAMS,SAK}, it is shown
that the matrix in the left-hand side of the first equation in
\Ref{19}, has a left eigenvector with the eigenvalue zero. (This
left eigenvector is ${\mathbf s}$.) So it does have a right
eigenvector with the eigenvalue zero as well. That is, there does
exist a vector ${\mathbf v}$ satisfying \Ref{19}. In fact, one can
even find a real vector ${\mathbf v}$ satisfying \Ref{19}. From
now on, ${\mathbf a}$ in \Ref{6} is assumed to satisfy \Ref{20}.
\section{Perturbations around autonomous systems}
Consider a system with the Hamiltonian $H$ as
\begin{equation}\label{21}
H=H^0+\delta H,
\end{equation}
where $H^0$ is the Hamiltonian of an autonomous system, and the
rates corresponding to $\delta H$ are small compared to those
corresponding to $H^0$. Our task is to investigate the evolution
one-point functions corresponding to $H$, using perturbation. As
$H$ is not necessarily autonomous, the evolution equation of the
one-point function may contain two-point functions as well.
However, the terms containing the two-point functions are
proportional to the rates corresponding to $\delta H$, and hence
are small. So, one can calculate the two-point function
corresponding to the unperturbed system, and use it in the
evolution equation of the one-point function of the perturbed
system, to obtain up-to-first-order evolution of the one-point
function of the perturbed system.
\subsection{The unperturbed solution}
Assuming that the initial condition is translationally-invariant,
it is seen that the one-point function is independent of the site,
and the two-point function depends on only the difference of the
sites' numbers. So, the evolution equation for the one-point
function of the unperturbed system is
\begin{equation}\label{22}
\frac{\d f^0}{\d t}=(\mu^0+\nu^0)f^0,
\end{equation}
where
\begin{equation}\label{23}
f^0:=\langle n_i\rangle^0,
\end{equation}
and
\begin{align}\label{24}
\mu^0=&{\mathbf s}\otimes{\mathbf a}\, H^0\,{\mathbf
u}\otimes{\mathbf v}+{\mathbf a}\otimes{\mathbf s}\, H^0\,{\mathbf
v}\otimes{\mathbf
u},\nonumber\\
\nu^0=&{\mathbf s}\otimes{\mathbf a}\, H^0\,{\mathbf
v}\otimes{\mathbf u}+{\mathbf a}\otimes{\mathbf s}\, H^0\,{\mathbf
u}\otimes{\mathbf v}.
\end{align}
The one-point function $f^0$, is easily seen to be

\begin{equation}\label{25}
f^0(t)=f^0(0)\exp[(\mu^0+\nu^0)t].
\end{equation}
Also, taking
\begin{equation}\label{26}
F_i^0:=\langle n_k\, n_{k+i}\rangle^0,
\end{equation}
(the two-point function of the unperturbed system) one arrives at
\begin{align}\label{27}
\frac{\d F_i^0}{\d t}&=\mu^0(F_{i-1}^0+F_{i+1}^0)+2\nu^0\,
F_i^0,\qquad
1<i<L\nonumber\\
\frac{\d F_1^0}{\d t}&=\mu^0\,
F_2^0+(\nu^0+\lambda^0)F_1^0+\rho^0\, f^0+\sigma^0,
\end{align}
where
\begin{align}\label{28}
\lambda^0:=&{\mathbf a}\otimes{\mathbf a}\, H^0\,{\mathbf
u}\otimes{\mathbf u},\nonumber\\
\rho^0:=&{\mathbf a}\otimes{\mathbf a}\, H^0\,({\mathbf
u}\otimes{\mathbf v}+{\mathbf v}\otimes{\mathbf u}),\nonumber\\
\sigma^0:=&{\mathbf a}\otimes{\mathbf a}\, H^0\,{\mathbf
v}\otimes{\mathbf v}.
\end{align}
It is seen that only five parameters enter the evolution equation
of the up-to-two-point functions, and all of these can be
expressed in terms of the matrix elements of
\begin{equation}\label{29}
\bar H^0:=H^0+\Pi\, H^0\,\Pi,
\end{equation}
where $\Pi$ is the permutation matrix. These parameters can be
rewritten as
\begin{align}\label{30}
\mu^0:=&{\mathbf s}\otimes{\mathbf a}\, {\bar H^0}\,{\mathbf
u}\otimes{\mathbf v}\nonumber\\
\nu^0:=&{\mathbf s}\otimes{\mathbf a}\, {\bar H^0}\,{\mathbf
v}\otimes{\mathbf u}\nonumber\\
\lambda^0:=&\frac{1}{2}{\mathbf a}\otimes{\mathbf a}\, {\bar H^0}\,{\mathbf u}\otimes{\mathbf u}\nonumber\\
\rho^0:=&{\mathbf a}\otimes{\mathbf a}\, {\bar H^0}\,{\mathbf
u}\otimes{\mathbf v}\nonumber\\
\sigma^0:=&\frac{1}{2}{\mathbf a}\otimes{\mathbf a}\, {\bar
H^0}\,{\mathbf v}\otimes{\mathbf v}.
\end{align}

Taking  a solution like
\begin{equation}\label{31}
F_{i}^0(t)=\sum_{E^0} F^0_{i\,E^0}(0)\exp(E^0\, t),
\end{equation}
it was shown in \cite{2point} that, the values of $E^0$
(energy-values) entering the two-point function are $0$,
$E^0_1:=\mu^0+\nu^0$, and any number in the interval
$I_0:=[2\nu^0-2|\mu^0|,2\nu^0 +2|\mu^0|]$, and possibly
\begin{equation}\label{32}
E^0_2:=\lambda^0+\nu^0+\frac{(\mu^0)^2}{\lambda^0-\nu^0}.
\end{equation}
$E^0_2$ is among the possible values of $E^0$, iff
\begin{equation}\label{33}
|\mu^0|\leq \lambda^0-\nu^0.
\end{equation}
The relation of $E^0_1$, $E^0_2$, and $I_0$, determines the
relaxation behavior of the two-point function (its dynamical
phase). Depending on the reaction rates several phases may occur
\cite{2point}:

\begin{itemize}
\item[\textbf{I)}] $E^0_1\in I_0$, and $E^0_2$ is not an energy. This is
the slower phase, and the longest relaxation time is
$[-2(\nu^0-\mu^0)]^{-1}$.

\item[\textbf{II)}] $E^0_1\in I_0$, and $E^0_2$ is an energy, in fact the
largest one. This is the slowest phase, and the longest relaxation
time is $\{-\nu^0-\lambda^0-[(\mu^0)^2/(\lambda^0-\nu^0)]\}^{-1}$.

\item[\textbf{III)}] $E^0_1> I_0$, and $E^0_2$ is not an energy. This is the
fastest phase, and the longest relaxation time is
$[-(\nu^0+\mu^0)]^{-1}$.

\item[\textbf{IV)}] $E^0_1> I_0$, $E^0_2$ is an energy, and $E^0_2<E^0_1$. This
is the fast phase, and the longest relaxation time is
$[-(\nu^0+\mu^0)]^{-1}$.

\item[\textbf{V)}]  $E^0_1> I_0$, $E^0_2$ is an energy, and $E^0_2>E^0_1$. This
is the slow phase, and the longest relaxation time is
$\{-\nu^0-\lambda^0-[(\mu^0)^2/(\lambda^0-\nu^0)]\}^{-1}$.
\end{itemize}
\subsection{The perturbed solution}
Now  consider the Hamiltonian $H$, defined through \Ref{21}, which
is not necessarily autonomous. Then defining
$\sideset{^{e}}{^{\alpha}{}_{\gamma\delta}}{\A}$ like \Ref{15} but
with $H$ instead of $H^0$, it is seen that the evolution equation
of the one-point function is
\begin{equation}\label{34}
\frac{\d \langle n_i \rangle}{\d t} =a_\alpha
\sideset{^1}{^{\alpha}{}_{\gamma\delta}}{\A}\langle N_i^\gamma
N_{i+1}^\delta\rangle+a_\alpha
\sideset{^2}{^{\alpha}{}_{\gamma\delta}}{\A}\langle N_{i-1}^\gamma
N_i^\delta\rangle.
\end{equation}
However, one cannot necessarily decompose
$\sideset{^{e}}{^{\alpha}{}_{\gamma\delta}}{\A}$ like \Ref{14}.
Assuming translational invariance of the initial conditions, one
arrives at
\begin{equation}\label{35}
\frac{\d f}{\d t} =a_\alpha \left(
\sideset{^1}{^{\alpha}{}_{\gamma\delta}}{\A}+
\sideset{^2}{^{\alpha}{}_{\gamma\delta}}{\A}\right) F_1^{\gamma
\delta},
\end{equation}
where
\begin{align}\label{36}
f:=&\langle n_i \rangle,\nonumber\\
F_1^{\gamma \delta}:=&\langle N_{i}^\gamma N_{i+1}^\delta\rangle.
\end{align}
As $\{\mathbf{u},\mathbf{v}\}$ is a basis, one can write $F_1$ in
terms of the tensor products of $\mathbf{u}$ and $\mathbf{v}$. The
corresponding coefficients can be found by multiplying the tensor
products of $\mathbf{a}$ and $\mathbf{s}$ by $F_1$. The result is
\begin{equation}\label{37}
F_1^{\gamma \delta}= F_1 u^\gamma u^\delta+f( u^\gamma v^\delta +
v^\gamma u^\delta) +  v^\gamma v^\delta.
\end{equation}
So,
\begin{equation}\label{38}
\frac{\d f}{\d t} =a_\alpha \left(
\sideset{^1}{^{\alpha}{}_{\gamma\delta}}{\A}+
\sideset{^2}{^{\alpha}{}_{\gamma\delta}}{\A}\right)\left[F_1
u^\gamma u^\delta+f( u^\gamma v^\delta + v^\gamma u^\delta) +
v^\gamma v^\delta\right].
\end{equation}
Defining
\begin{equation}\label{39}
\sideset{^e}{^{\alpha}{}_{\gamma\delta}}\B :=
\sideset{^e}{^{\alpha}{}_{\gamma\delta}}\A-\sideset{^e}{^{0\alpha}{}_{\gamma\delta}}\A,
\end{equation}
\Ref{38} recasts to
\begin{align}\label{40}
\frac{\d f}{\d t} =&a_\alpha
(\sideset{^1_1}{^{0\alpha}{}_\gamma}\A\,s_\delta+
\sideset{^1_2}{^{0\alpha}{}_\delta}\A\,s_\gamma
+\sideset{^2_1}{^{0\alpha}{}_\gamma}\A\,s_\delta+
\sideset{^2_2}{^{0\alpha}{}_\delta}\A\,s_\gamma+
\sideset{^1}{^{\alpha}{}_{\gamma\delta}}\B+
\sideset{^2}{^{\alpha}{}_{\gamma\delta}}\B )
\nonumber \\
&\times \left[F_1 u^\gamma u^\delta+f( u^\gamma v^\delta +
v^\gamma u^\delta) + v^\gamma
v^\delta\right]\nonumber\\
=&(\mu^0+\nu^0)f+(\sideset{^1}{{}_{\gamma\delta}}\BB+
\sideset{^2}{{}_{\gamma\delta}}\BB) \left[F_1 u^\gamma u^\delta+f(
u^\gamma v^\delta + v^\gamma u^\delta) + v^\gamma v^\delta\right],
\end{align}
where
\begin{equation}\label{41}
\sideset{^e}{{}_{\gamma\delta}}\BB:=
a_\alpha\sideset{^e}{^{\alpha}{}_{\gamma\delta}}\B.
\end{equation}
As expected, the coefficients of $F_1$ in the right-hand side are
small (first order in terms of the perturbation $\delta H$). So,
one can use the zeroth-order value of $F_1$ in the right-hand
side, to obtain the first-order value of $f$.

From \Ref{25} and \Ref{27}, it is seen that if $E^0_1>I_0$, then
one can write
\begin{equation}\label{42}
F^0_1=\hat F^0_1+\frac{\rho^0}{\mu^0-\lambda^0-\mu^0\,z}f^0,
\end{equation}
where $z$ satisfies
\begin{equation}\label{43}
\mu^0+\nu^0=\mu^0(z+z^{-1})+2\nu^0,
\end{equation}
and its modulus is less than 1, and $\hat F^0_1$ is like \Ref{31},
but without a term corresponding to the energy-value
$E^0_1=\mu^0+\nu^0$. So, one can write \Ref{40} like
\begin{align}\label{44}
\frac{\d f}{\d t}
=&(\mu^0+\nu^0)f+(\sideset{^1}{{}_{\gamma\delta}}\BB+
\sideset{^2}{{}_{\gamma\delta}}\BB)\left(\frac{\rho^0}{\mu^0
-\lambda^0-\mu^0\,z} u^\gamma u^\delta+u^\gamma v^\delta +
v^\gamma u^\delta\right)f\nonumber \\
&+(\sideset{^1}{{}_{\gamma\delta}}\BB+
\sideset{^2}{{}_{\gamma\delta}}\BB)(\hat F_1 u^\gamma u^\delta+
v^\gamma v^\delta).
\end{align}
This means that the energy-values entering $f$, are those entering
$\hat F^0_1$, and
\begin{equation}\label{45}
E_1:=\mu+\nu+\frac{\rho^0}{\mu^0 -\lambda^0-\mu^0\,z}\delta\theta,
\end{equation}
where
\begin{align}\label{46}
\mu:=&\mu^0 +\delta \mu,\nonumber\\
\nu:=&\nu^0 +\delta \nu,\nonumber\\
\delta \mu:=&{\mathbf s}\otimes{\mathbf a}\, { \delta \bar
H}\,{\mathbf
u}\otimes{\mathbf v}\nonumber\\
\delta \nu:=&{\mathbf s}\otimes{\mathbf a}\, { \delta \bar
H}\,{\mathbf
v}\otimes{\mathbf u}\nonumber\\
\delta \theta :=&{\mathbf a}\otimes{\mathbf s}\, {\delta \bar
H}\,{\mathbf u}\otimes{\mathbf u},
\end{align}
and
\begin{equation}\label{47}
\delta \bar H:=\delta H +\Pi\, \delta H\,\Pi.
\end{equation}
This shows that if $E^0_1$ is the largest nonzero energy-value
entering $F^0_1$, then $E_1$ is the largest nonzero energy-value
entering $f$. Otherwise, the largest nonzero energy-value entering
$f$ is the largest nonzero energy-value entering $F^0_1$. So the
relaxation behavior of $f$, can be deduced from that of $F^0_1$ as
follows:
\begin{itemize}
\item[\textbf{I)}] In this phase the largest nonzero energy-value of $f$ is
$2(\nu^0-\mu^0)$, and the perturbation causes a discontinuous
change of the largest nonzero energy-value, from
$E^0_1=\mu^0+\nu^0$ to $2(\nu^0-\mu^0)$.
\item[\textbf{II,V)}] In this phase the largest nonzero energy-value of $f$ is
$E^0_2$, and the perturbation causes a discontinuous change of the
largest nonzero energy-value, from $E^0_1=\mu^0+\nu^0$ to $E^0_2$.
\item[\textbf{III, IV)}] In this phase the largest nonzero energy-value of
$f$ is $E_1$, and the perturbation causes a continuous change of
the largest nonzero energy-value, from $E^0_1$ to $E_1$.
\end{itemize}
It is seen that the perturbation causes two different changes in
the relaxation behavior of $f$ (the one-point function). In the
regions \textbf{I}, \textbf{II}, and \textbf{V}, the perturbation
causes a discontinuous change in the relaxation behavior, which
means that the autonomous system is unstable with respect to
perturbations. In the regions \textbf{III} and \textbf{IV},
however, the relaxation behavior of $f$ is continuous with respect
to the perturbations, which means that the autonomous system is
stable with respect to the perturbations, at least as long as
first-order perturbations of the one-point function are
considered. Mentioning one other thing is also in order: $F^0_1$
enters the evolution equation of $f$, iff $\delta\theta\ne 0$. If
$\delta\theta=0$, then $f$ would contain only one (nonzero)
energy-value, which is the one expected from changing one
autonomous system to another. (Only $\mu^0$ is replaced by $\mu$
and $\nu^0$ is replaced by $\nu$.)

A real autonomous system, would in fact be only approximately
autonomous. This means that there are always perturbations around
the autonomous system. The above argument shows that only
autonomous systems in the regions \textbf{III} and \textbf{IV} can
be effectively autonomous. The parameter space corresponding to
the effectively-autonomous systems is
\begin{equation}\label{48}
-\frac{\mu^0}{\nu^0}>-\frac{1}{3},\quad -\frac{\lambda^0}{\nu^0}>
\frac{(-\mu^0/\nu^0)-1+\sqrt{[1+3(-\mu^0/\nu^0)][1-(-\mu^0/\nu^0)]}}{2}.
\end{equation}
Figure 1 shows the regions corresponding to the autonomous system.
(This is identical to figure 1 in \cite{2point}.)

\section{An example}
Consider an Hamiltonian $H^0$ corresponding to an autonomous
system:
\begin{equation}\label{49}
H^0=\frac{1}{4}
\begin{pmatrix}
-3+3\omega&\omega&\omega&1-\omega\\
1-\omega&-3\omega&\omega&1-\omega\\
1-\omega&\omega&-3\omega&1-\omega\\
1-\omega&\omega&\omega&-3+3\omega\\
\end{pmatrix}
+r
\begin{pmatrix}
-1&0&0&0\\
0&0&0&1\\
0&0&0&1\\
1&0&0&-2\\
\end{pmatrix}.
\end{equation}
The reactions of the corresponding system are
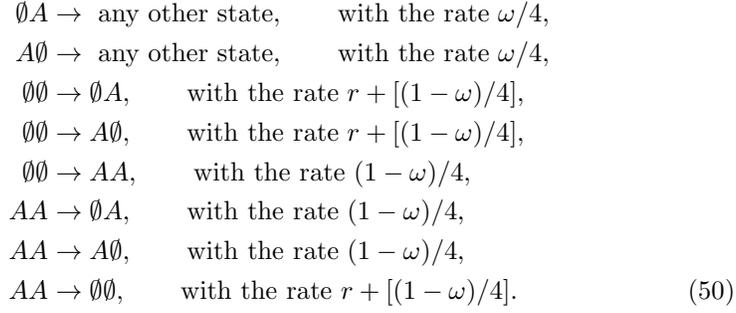
\begin{align}\label{50}
\emptyset A&\to\hbox{ any other state},\qquad\hbox{with the rate
$\omega/4$},\nonumber\\
A\emptyset&\to\hbox{ any other state},\qquad\hbox{with the rate
$\omega/4$},\nonumber\\
\emptyset\emptyset&\to\emptyset A,\qquad\hbox{with the rate
$r+[(1-\omega)/4]$},\nonumber\\
\emptyset\emptyset&\to A\emptyset,\qquad\hbox{with the rate
$r+[(1-\omega)/4]$},\nonumber\\
\emptyset\emptyset&\to AA,\qquad\hbox{with the
rate $(1-\omega)/4$},\nonumber\\
AA&\to\emptyset A,\qquad\hbox{with the rate
$(1-\omega)/4$},\nonumber\\
AA&\to A\emptyset,\qquad\hbox{with the rate
$(1-\omega)/4$},\nonumber\\
AA&\to\emptyset\emptyset,\qquad\hbox{with the rate
$r+[(1-\omega)/4]$}.
\end{align}
For this Hamiltonian, one has
\begin{align}\label{51}
\mathbf{v}=\frac{1}{2}
  \begin{pmatrix}
    1\\
    1
  \end{pmatrix},\nonumber\\
\mathbf{a}=\frac{1}{2}
  \begin{pmatrix}
    1&1
  \end{pmatrix},
\end{align}
and
\begin{align}\label{52}
\mu^0=&-1-2r+2\omega,\nonumber\\
\nu^0=&-1-2r,\nonumber\\
\lambda^0=&-\frac{1}{2}-r,\nonumber\\
\rho^0=&r.
\end{align}
For the perturbation, consider the Hamiltonian
\begin{equation}\label{53}
\delta H=\varepsilon
\begin{pmatrix}
-1&0&0&0\\
0&0&0&0\\
0&0&0&0\\
1&0&0&0\\
\end{pmatrix}.
\end{equation}
It is easy to see that this Hamiltonian does not correspond to an
autonomous system. This Hamiltonian only increases the rate of the
reaction $AA\to\emptyset\emptyset$ by $\varepsilon$. Using this
Hamiltonian, it is seen that
\begin{align}\label{54}
\delta\mu&=-\varepsilon,\nonumber\\
\delta\nu&=-\varepsilon,\nonumber\\
\delta\theta&=-2\varepsilon.
\end{align}
From \Ref{52}, it is seen that
\begin{align}\label{55}
-\frac{\mu^0}{\nu^0}&=-1+2\frac{\omega}{1+2r},\nonumber\\
-\frac{\lambda^0}{\nu^0}&=-\frac{1}{2}.
\end{align}
Comparing this with \Ref{48}, it is seen that the system is
effectively autonomous, iff
\begin{equation}\label{56}
\omega>\frac{5-\sqrt{5}}{8}(1+2 r).
\end{equation}

As a special case of the above example, let us put $\omega=1$. In
this case, for the nonperturbed system we have the following
reactions.
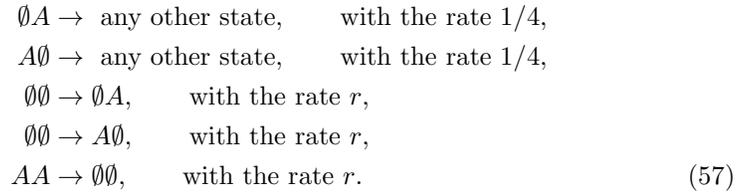
\begin{align}\label{57}
\emptyset A&\to\hbox{ any other state},\qquad\hbox{with the rate
$1/4$},\nonumber\\
A\emptyset&\to\hbox{ any other state},\qquad\hbox{with the rate
$1/4$},\nonumber\\
\emptyset\emptyset&\to\emptyset A,\qquad\hbox{with the rate
$r$},\nonumber\\
\emptyset\emptyset&\to A\emptyset,\qquad\hbox{with the rate
$r$},\nonumber\\
AA&\to\emptyset\emptyset,\qquad\hbox{with the rate $r$}.
\end{align}
In this case, the system is effectively autonomous iff
\begin{equation}\label{58}
r<r_0:=\frac{1}{2}+\frac{1}{\sqrt{5}}.
\end{equation}
It is seen that changing the value of $r$ from $0$ to $+\infty$,
the system starts from phase \textbf{III}, passes through the
phases \textbf{IV}, \textbf{V}, and \textbf{II}, and finally
reaches the phase \textbf{I}. At $r=r_0$, the system goes from the
phase \textbf{IV} to the phase \textbf{V}, which means that the
system is no longer effectively autonomous.

\end{document}